\nonstopmode
\documentclass[a4paper,twoside]{article}
\usepackage[authoryear]{natbib}
\bibpunct{(}{)}{;}{a}{}{,}
\usepackage{aas_macros}
\usepackage{fix2col}
\usepackage{graphicx}
\usepackage{times}
\usepackage{url}
\date{} 
\title{\large\bf\flushleft The ATCA Seeing Monitor}
\author{\parbox{\textwidth}{\flushleft
\vspace{-0.5cm}
{\it Enno Middelberg, Robert J. Sault, Michael J. Kesteven}\\
\vspace{0.4cm}
{\small Australia Telescope National Facility, PO Box 76, Epping NSW 1710, Australia}\\
{\small Email: enno.middelberg@csiro.au}}}
\begin{document}
\twocolumn[
\begin{changemargin}{.8cm}{.5cm}
\begin{minipage}{.9\textwidth}
\vspace{-1cm}
\maketitle
%
%
\small{\bf Abstract:}

We present the design of, and a first analysis of data from, the
atmospheric seeing monitor at the Australia Telescope Compact Array
(ATCA). The seeing monitor has been operational almost continuously
since May 2004 and every 10\,min delivers a measurement of the
atmospheric phase stability at the observatory. Its measurements can
be used by observers to help deciding whether it is worth carrying out
observations at millimetre wavelengths or whether a longer-wavelength backup
project should be observed. We present a statistical analysis of the
data recorded since September 2004 to characterize the annual variations in
atmospheric path length fluctuations. Our analysis shows that in terms
of phase stability, nights in spring, summer, and autumn are as good
as, or better than, days in winter. We also find that the
data imply that the turbulence in the lower few hundred metres
of the atmosphere is predominantly responsible for the atmospheric seeing.


\medskip
\medskip
\end{minipage}
\end{changemargin}
]
\small

\section{Introduction}

The dominant source of phase errors in interferometric radio
observations at frequencies above 5\,GHz are fluctuations in the
tropospheric water vapour content along the line of sight of the
individual interferometer elements. The effect of the fluctuations
scales linearly with frequency, making observations at frequencies of
tens of GHz (at wavelengths less than about one cm) particularly prone
to atmospheric phase changes. For a radio interferometer operating at
frequencies of up to 115\,GHz such as the Australia Telescope Compact
Array (ATCA) it is therefore desirable to identify the periods within
a year that are suitable for observations at these high frequencies. A
well-tested means of characterizing the phase stability above a radio
observatory are two-element interferometers observing unmodulated
beacons of geostationary satellites (e.g. \citealt{Radford1996,
Hiriart2002}).

\subsection{Atmospheric seeing in the radio regime}

A excellent introduction to radio seeing can be found in
\citealt{Thompson2001}. It is generally assumed that the water vapour
distribution in the troposphere can be described as follows. Water
vapour does not mix well with air because the air temperature is close
to the condensation point of water. Its distribution can be
approximated as nested parcels of air with a variety of water vapour
densities. Kinetic energy is transferred from larger parcels to
smaller and smaller parcels until it is dissipated by viscous
friction, a situation known as Kolmogorov turbulence. A sketch of the
situation can be found in Figure~II in \cite{Masson1994}. The rms
phase fluctuations ($\sigma_\phi$), measured over a long period, are
expected to exhibit a power law relationship with baseline length,
$b$:

\begin{equation}
\sigma_\phi^2 \propto b^{\beta}.
\label{eq:bk}
\end{equation}

Generally the turbulence is considered to be confined to a specific
layer.  When the layer is much thinner than the interferometer
spacing, two-dimensional turbulence theory is applicable, which
suggests a power law exponent of $\beta/2 = 1/3$ for the rms
phase. When the layer is deep compared with the interferometer
spacing, three-dimensional theory is appropriate, and a power law
exponent of $\beta/2 = 5/6$ is expected.

Furthermore, it is assumed that the mixture of air and water vapour is
``frozen'' in the atmosphere and is transported relative to the ground
at a particular speed, $v_{\rm s}$. This wind speed aloft couples the
temporal and spatial behaviour of the phase screen, and $v_{\rm s}$
therefore is called the phase screen speed.

\section{Design of the seeing monitor}

Here we report on the development of a seeing monitor at the ATCA,
and present an analysis of some data.

The ATCA seeing monitor is an interferometer on a 230\,m east-west
baseline and tracks the 30.48\,GHz beacon on the geostationary
communications satellite, OPTUS-B3, at an elevation of
$60^\circ$. Each element of the interferometer consists of a 1.8\,m
prime-focus paraboloid equipped with an uncooled low-noise amplifier
(LNA).  The FWHM of the reflectors' main lobes is 0.4$^\circ$. This is
large compared to the satellite's diurnal motion on the sky
(0.1$^\circ$), and hence the reflectors are stationary and do not need
to track.

The power arriving at the feed horns is approximately
$3\cdot10^{-13}$\,W, or -125\,dBW. The signals are received and
amplified in the primary foci, then transported to a screened
temperature-controlled enclosure next to the antenna, where the
signals are down converted in two steps to a frequency of
240\,MHz. They are then sent to the ATCA's screened room via optical
fibres, where they are mixed down to a frequency of 50\,kHz, and
analysed.

The design of the system was challenged by the poor signal-to-noise
ratio, and the drifting beacon frequency due to the satellite's radial
diurnal motion: this motion makes the beacon frequency drift through
approximately 5\,kHz (peak-to-peak) each day.

The problem of the changing frequency was addressed by coupling the
local oscillator frequency to the satellite beacon. This is
accomplished with a Stanford Research Systems SR510 lock-in amplifier
with a very narrow-band (2\,Hz) notch filter, which tracks the
satellite beacon frequency automatically.

The signal-to-noise ratio was improved by generating a clean signal
coupled to the notch-filtered frequency and using this signal as a
reference to which the other antenna's signal is to be compared. This
is realised with a Stanford Research Systems SR830 lock-in
amplifier. The phase measurement is then carried out with a second
SR510.

\begin{figure*}[htpb!]
\centering
\includegraphics[width=\linewidth]{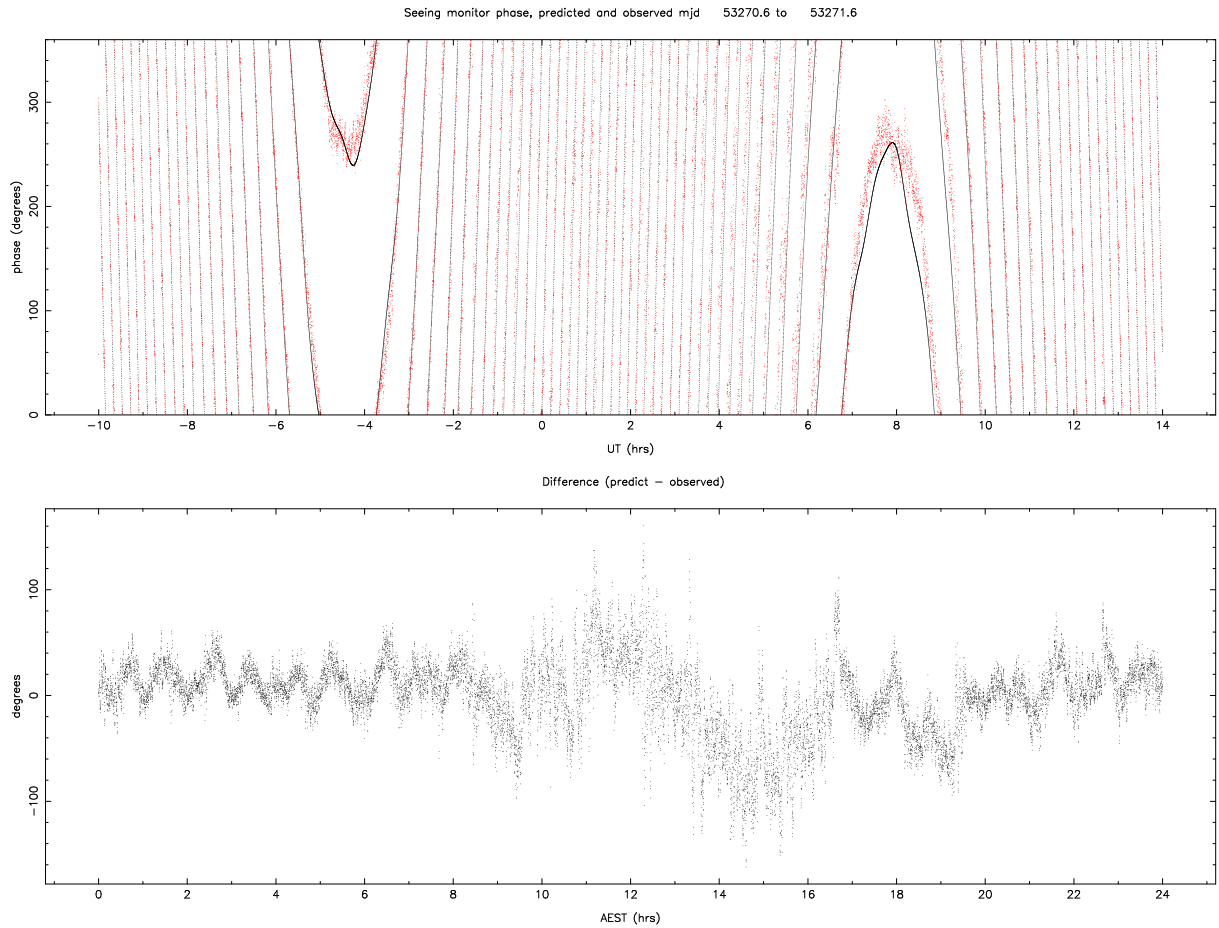}
\caption{Comparison of the predicted and measured satellite phase on 23
September 2004. {\it Top panel:} The predicted (solid line) and
measured (dots) satellite phase. The x axis shows time in UT hours and
the y axis the phase in degrees. {\it Bottom panel:} The difference
between the prediction and the measurements. The x axis shows time in
AEST hours and the y axis the residual phase in degrees. The periodic
structure in the residuals is due to the cycling of the air
conditioning in the receiver enclosure. The relatively quiet sections
before 9\,h AEST and after 17\,h AEST emphasize the increase in path
length fluctuations during the day (cf. Fig.~\ref{fig:time}).}
\label{fig:satphase}
\end{figure*}

In addition to the atmosphere, the interferometer phase is affected by
two other phenomena: both the small motion of the satellite and thermal
drifts in the receiving system introduce extra phase responses.
Figure~\ref{fig:satphase} shows a comparison between the measured phase
and that predicted purely from the satellite's motion.  The residuals
after subtracting off this component clearly show both the atmospheric
and thermal components.

\section{Observations and data analysis}

\subsection{Operation of the seeing monitor}

The integration times of the seeing monitor's lock-in amplifiers are
set to 100\,ms, however, the phase is read and archived by the
observatory monitor system only every 5\,s. The standard processing of
the data are as follows: the difference between two successive phase
measurements is found, the standard deviation of this difference is
computed and the measurement is converted into a path length using
Eq.~(\ref{eq:rms1}) before archiving. By taking the standard deviation
(as distinct from rms) of differences, any response to a linear drift
within the averaging period is eliminated. Thus the effect of the
satellite motion and thermal phase drifts are removed.  This standard
deviation is computed over 10\,min of data.  As 10\,min is long
compared to the wind-crossing time over the seeing monitor baseline
(the crossing time is usually no more than one minute), this is a
statistically adequate period for estimating the rms phase.

\begin{figure}[htpb]
\centering
\includegraphics[width=\linewidth]{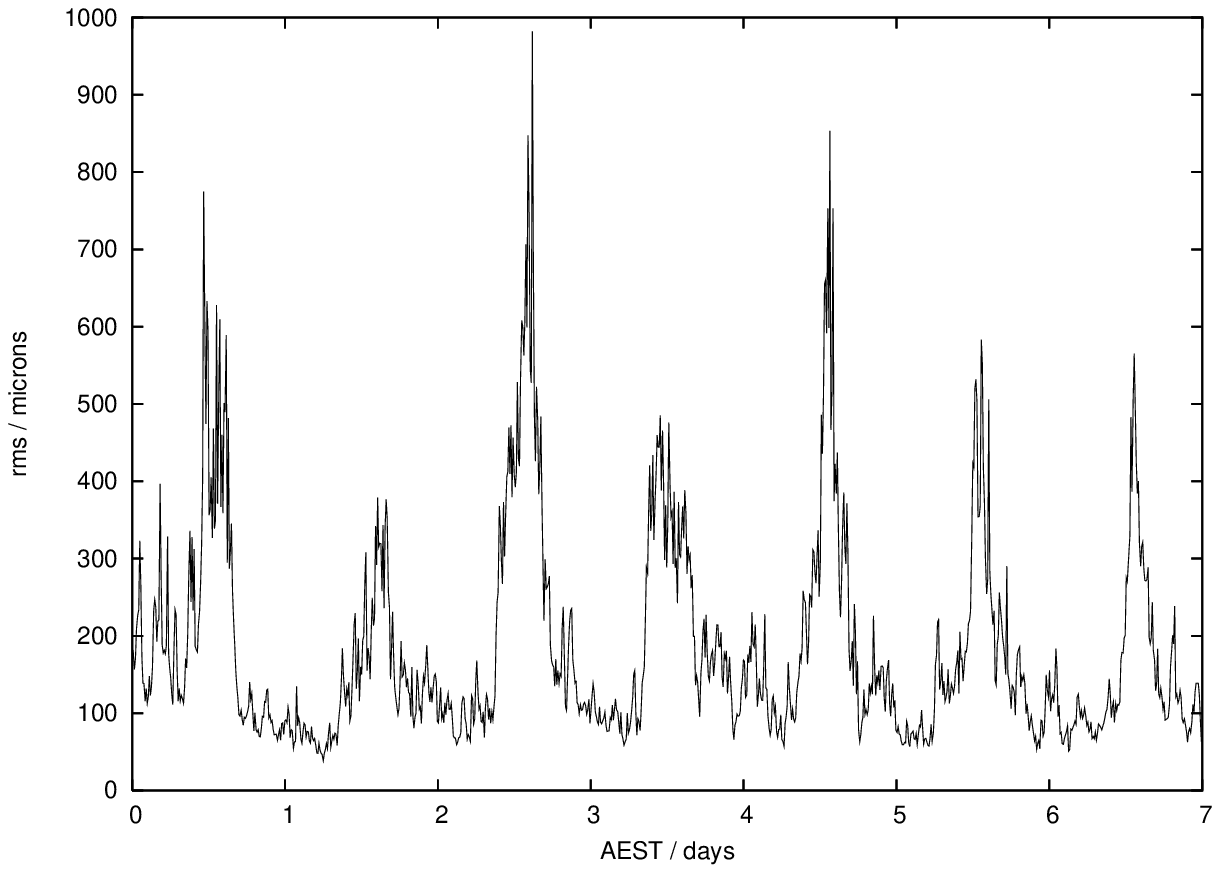}
\caption{Example of seven days of seeing monitor data taken between 14
and 21 May 2005. The labelled major tic marks correspond to 00:00\,h
AEST. The diurnal variations can be clearly seen, with relatively
quiet nights and pronounced peaks around noon and in the early
afternoon. Note that these values represent the seeing at $60^\circ$
elevation on a 230\,m baseline.}
\label{fig:example}
\end{figure}

The real-time measurements and plots from the archive are accessible
to staff and observers.  As an example, the path length fluctuation
from a seven day period are reproduced in Fig.~\ref{fig:example}.

\subsection{Our analysis}

The analysis presented in this paper is more detailed than the
analysis carried out routinely at the observatory and yields more
insights into the properties of the atmosphere.

We have converted the raw seeing monitor phases into path length
measurements in microns and divided them into sections of 30\,min. A
second-order polynomial was subtracted from the measurements to
eliminate changes due to satellite motion and thermal drifts in this
interval. We have calculated the rms of the residuals, which is a
measure of the radio seeing in this interval.

\begin{figure*}[htpb!]
\centering
\includegraphics[width=\linewidth]{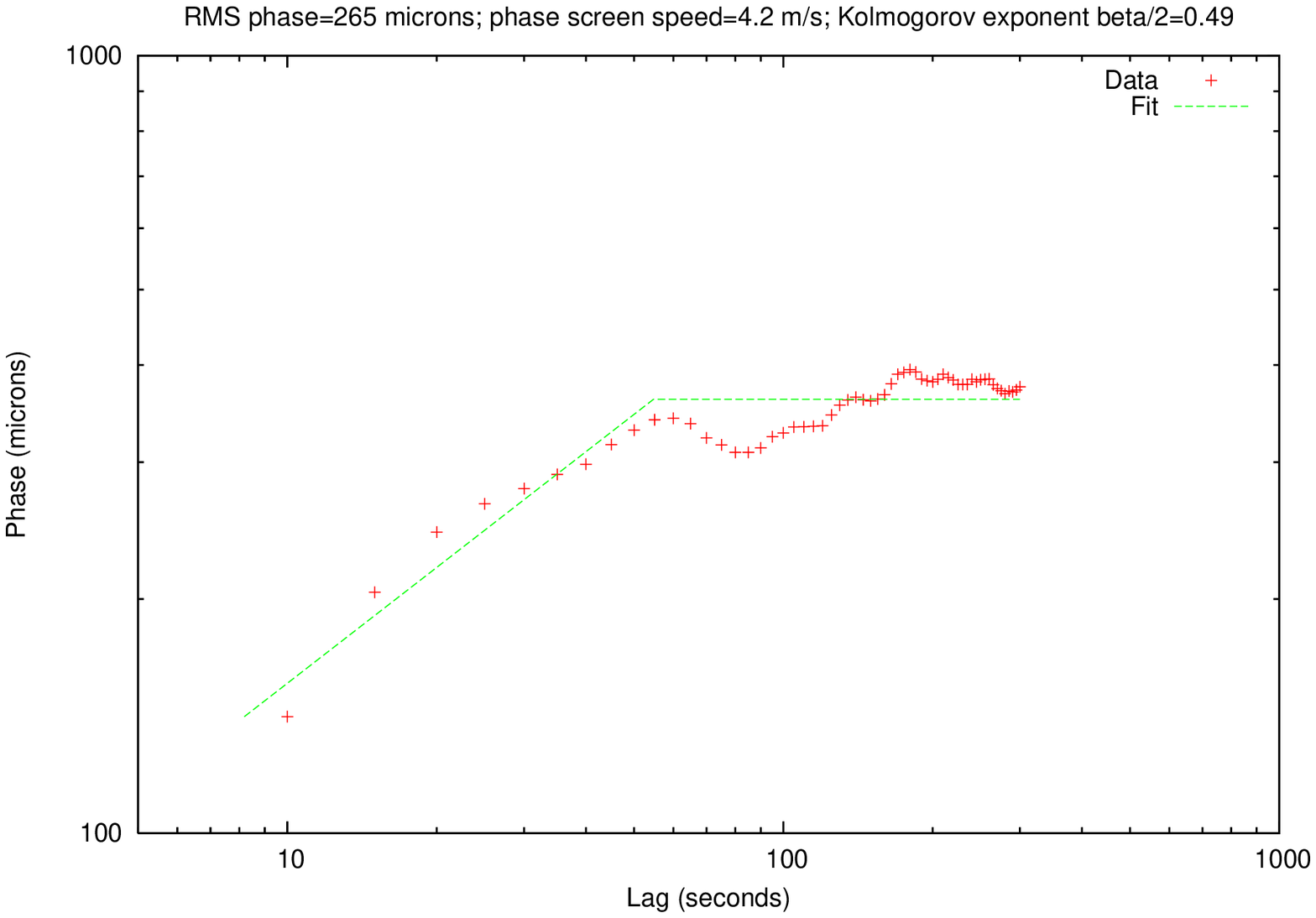}
\caption{Typical lag phase structure function and a fit
to these data. The data are taken on 16 June 2006.}
\label{fig:structurefunc}
\end{figure*}

Furthermore, using these 30\,min intervals of data, we have followed
an approach using lag structure functions (see \citealt{Holdaway1995}) to
determine the phase screen speed and Kolmogorov exponent.  In a
logarithmic diagram, the structure function of the path length
fluctuations increases linearly with lag, until it flattens out after
a characteristic time. The slope of the linear increase is a measure
of the Kolmogorov exponent, $\beta/2$, the time after which saturation
occurs is characteristic of the speed of the phase screen, and the
level at which saturation occurs is expected to be $\sqrt{2}$ times
the rms path length fluctuations. We have measured these quantities in
each 30\,min interval by fitting linear functions to the two parts of
the structure functions (Fig~\ref{fig:structurefunc}).

The simple rms of the data, and the fit to the saturation level of the
lag structure function give two methods to estimate $\sigma_\phi$.
The average of the ratio of these two estimates, after correcting for
the factor of $\sqrt{2}$, is $0.995\pm0.098$, which is a near-perfect
agreement.

It is common to normalise the rms of the path length fluctuations to an
interferometer baseline of 1\,km, pointing towards the zenith. We
therefore have multiplied all rms measurements with $\sqrt{3}/2$ to
convert the measurements to zenith values, and by $(1000/230)^{\beta/2}$,
to scale the measurements from the seeing monitor's 230\,m baseline to
1\,km, where $\beta/2$ is the Kolmogorov exponent derived from the
structure function in each interval.

The data presented here were observed in the period of 16 September
2004 to 25 May 2006. There are only two noticeable periods where the
seeing monitor was either not observing or the measurements were below
$20\,\mu$m, indicating a malfunction. These periods are 7 October 2004
to 19 October 2004 (12.2\,days); and 12 January 2005 to 14 April 2005
(92.3\,days). There are 9 shorter (one to five days) periods and many
still shorter periods (typically a few hours) without data.  The
shortest outages tend to happen preferentially in summer and are
associated with storms (the seeing monitor fails when the sky becomes
opaque to the 30~GHz beacon, and under some situations when mains
power had been lost). We have performed some analysis to convince
ourselves that these short gaps in the data do not significantly
affect or bias our conclusions below.

The effect of the longer gaps is more difficult to estimate. The
missing data essentially mean that some times of the year were sampled
only once, and so are more prone to rare weather situations that may
bias the characteristics of that period.

\section{Impact on the scheduling of millimetre observations}

\subsection{Path length fluctuations as a function of time of day and
time of year}

\begin{figure}[htpb]
\centering
\includegraphics[width=\linewidth]{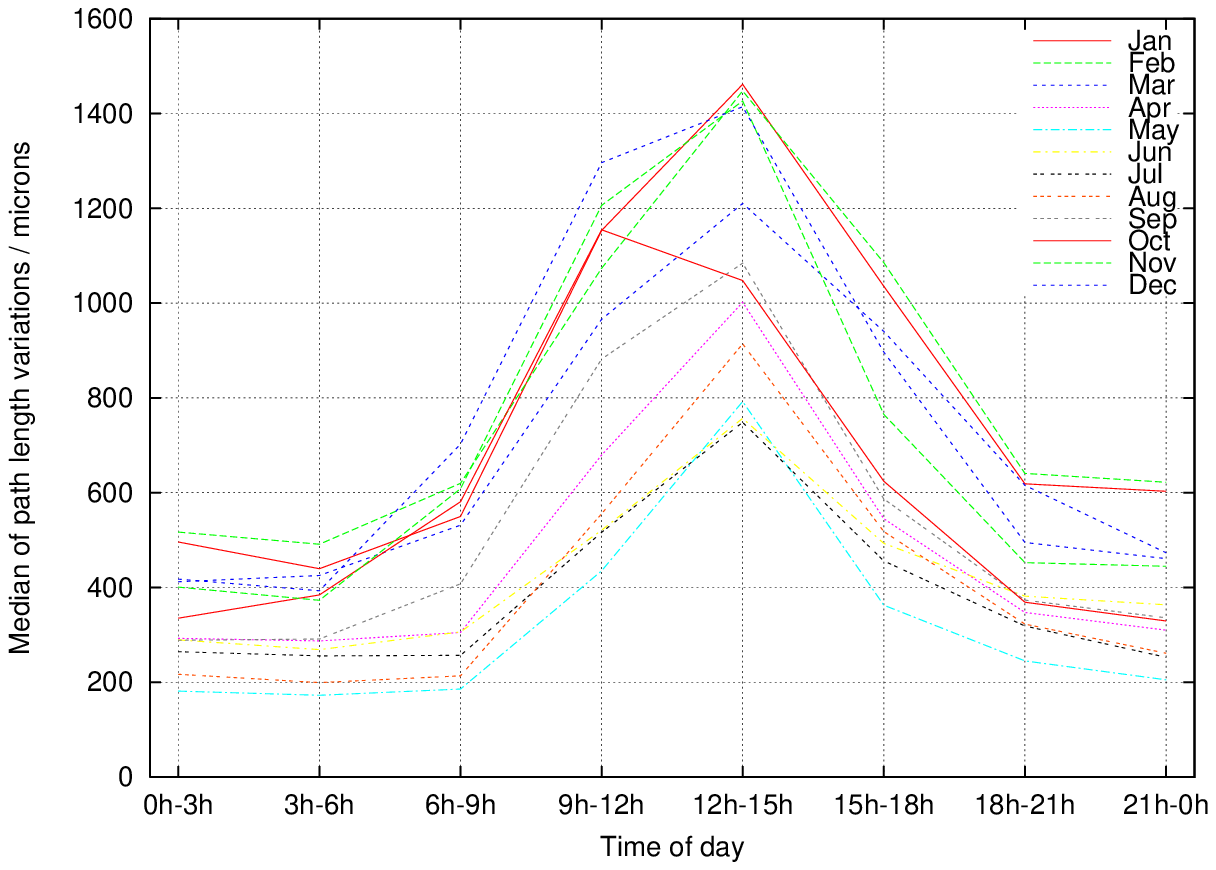}
\caption{Rms of the path length variations as a function of time of day.}
\label{fig:time}
\end{figure}

We have sorted the data by month and time of day (local time,
AEST). The data observed within each month were divided into eight
groups, each of which represents 3\,h of a day in that particular
month (00:00 to 03:00, then 03:00 to 06:00, and so on). The number of
measurements in each of these groups is of the order of 220 (covering
the range of 114 to 336 measurements). Measurements of less than
$20\,\mu$m have been deemed instrumental errors and have been
ignored. The medians of the rms phase these groups are plotted in
Figure~\ref{fig:time}. It shows that throughout the year, the highest
rms values occur in the period between 12:00 and 15:00 AEST. The only
exception is October, where the maximum occurs in the period
09:00-12:00. The plot also reveals that in any one month at night
(between 21:00 and 06:00 AEST), the rms is lower than 650\,$\mu$m,
which is lower than the maximum rms in any month. This is interesting
because it suggests that from the perspective of atmospheric seeing,
summer nights are as good as, or better than, winter days. Hence it
may be feasible to extend the period in which millimetre projects are
scheduled. However, from Figure~\ref{fig:time} it is not easy to tell
by how much summer nights are better. We quantify this in the
following section.

\subsection{Quantitative analysis of path length fluctuations}

At the ATCA, the ``millimetre season'', i.e., the time of year during
which observations at millimetre wavelengths (currently frequencies of
86\,GHz and higher) are scheduled, is early May to August. Observing
time in this period is in high demand, and to ensure the best possible
usage of observing time it is desirable to know to what extent the
millimetre season can be expanded into the shoulder seasons. The
seeing monitor measurements from the millimetre season should
therefore be compared to measurements taken throughout the rest of the
year.

Furthermore, we deem typical daytime conditions to occur between
09:00 and 12:00, and between 15:00 and 18:00. The rms is
remarkably worse between 12:00 and 15:00 in all months, and not
representative of typical daytime conditions. Nighttime conditions
appear to occur between 21:00 and 06:00, where the path length
fluctuations are relatively similar. Hence, we compare measurements
taken in the millimetre season months (May to August) between 09:00 and
12:00, and between 15:00 and 18:00, to measurements taken in
all other months between 21:00 and 06:00.

A measure of what fraction of time in these periods is useful can be
obtained by estimating the amount of decorrelation arising from
atmospheric seeing. While the amount of decorrelation observers are
prepared to accept may vary, 10\,\% appears to be a reasonable amount
which would allow one to carry out many experiments. From expression
(\ref{eq:rms3}), the corresponding rms of the path length fluctuations
follows to $255\,\mu$m for a wavelength of 3.49\,mm (a frequency of
86\,GHz). This needs to be scaled to a 1\,km baseline to be comparable
to our measurements, which requires a typical baseline length used in
millimetre observations, $d$, and an estimate for the Kolmogorov
exponent. We assume $d=150\,{\rm m}$, and use the median Kolmogorov
exponent of all measurements taken in the months of May to August (the
millimetre season), which is $\beta/2=0.57$. The rms of the path
length fluctuations then scales to $255\,\mu{\rm m}\times
(\frac{1000\,{\rm m}}{150\,{\rm m}})^{0.57}=752\,\mu{\rm m}$.

\begin{figure}[htpb]
\centering
\includegraphics[width=\linewidth]{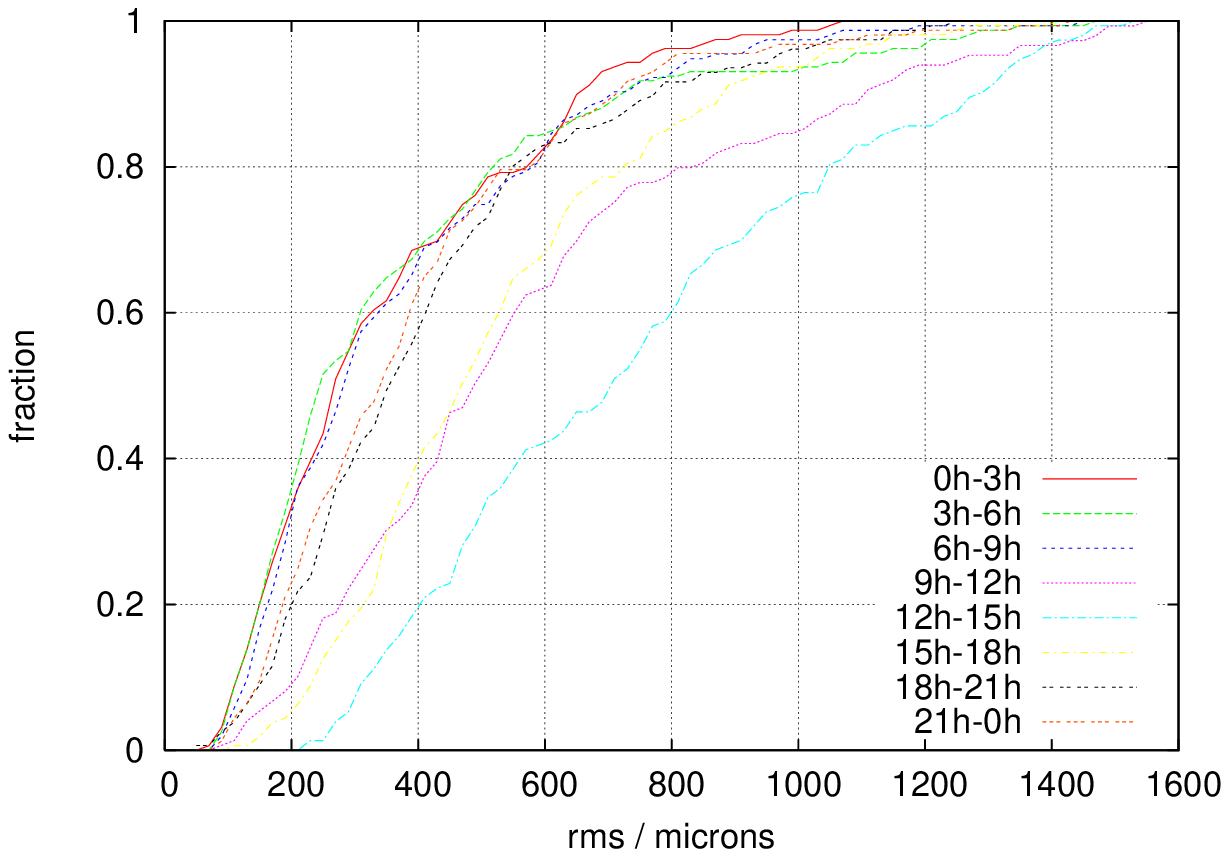}
\includegraphics[width=\linewidth]{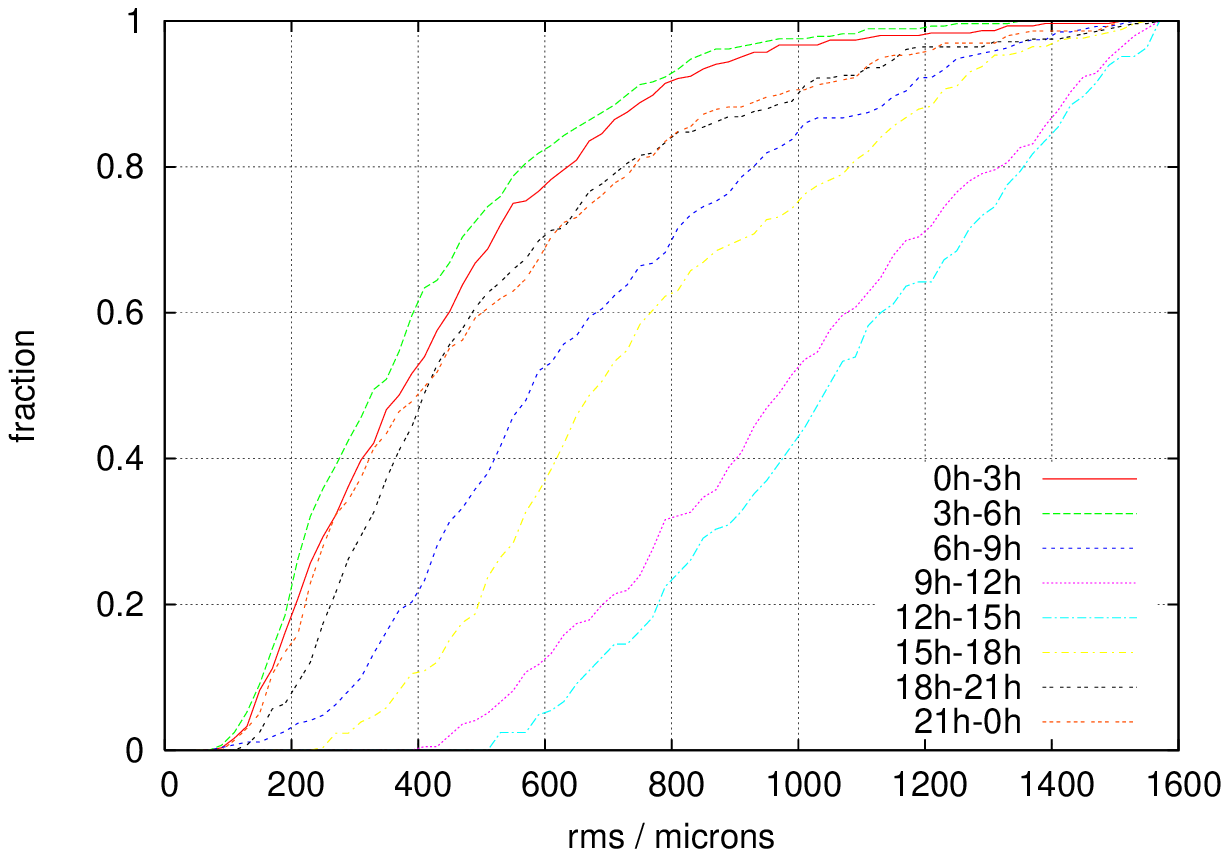}
\includegraphics[width=\linewidth]{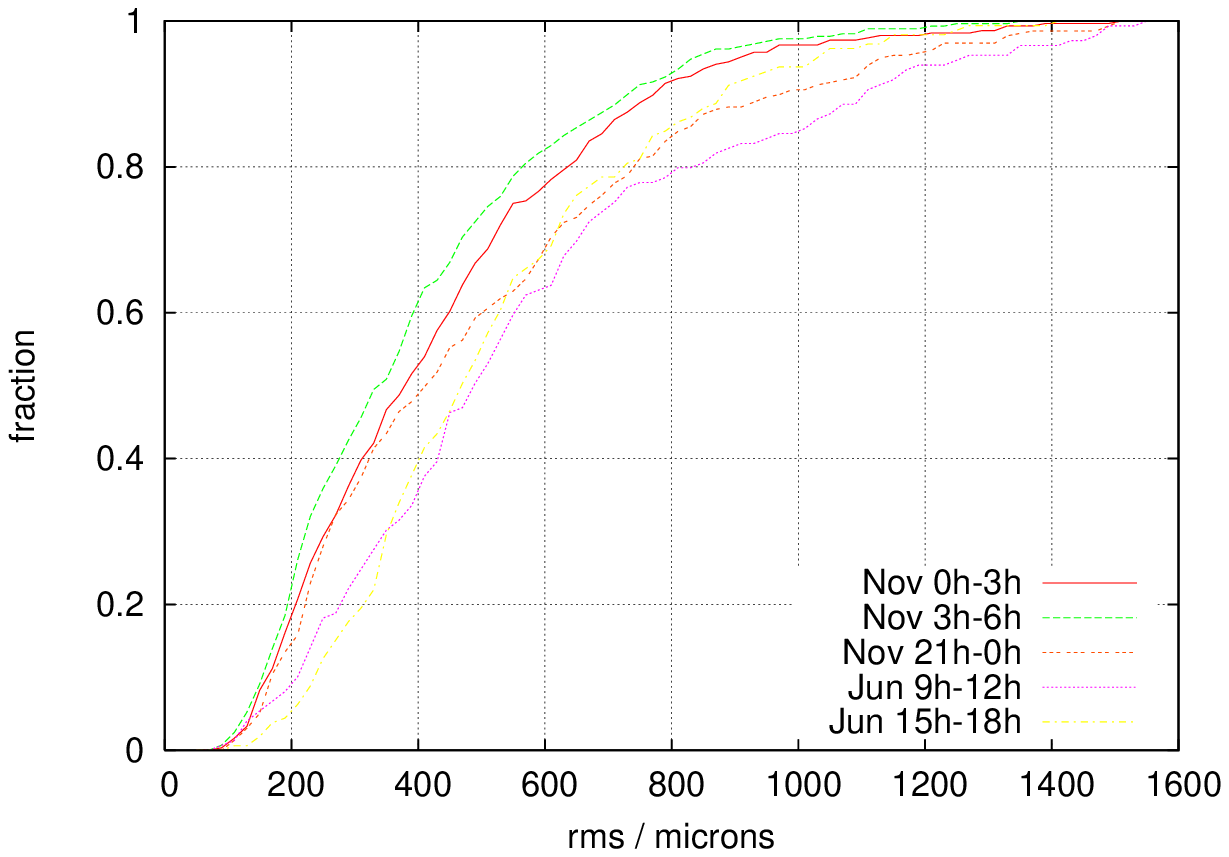}
\caption{{\it Top Panel:} Cumulative histogram for each 3\,h bin of the
day in June. {\it Middle Panel:} Cumulative histogram for each bin
of the day in November. Comparing the two uppermost panels illustrates
the generally better seeing conditions in winter. {\it Bottom panel:}
Comparison of typical daytime conditions in June and typical nighttime
conditions in November. Summer nights appear to be slightly better
than winter days in this diagram.}
\label{fig:cumhist}
\end{figure}

For each month, we have constructed cumulative histograms of the data
in each 3\,h period of the day, yielding the fraction of measurements
below a given rms value. Samples are reproduced in
Figure~\ref{fig:cumhist}. For example, in the top panel, representing
all measurements of June, the ``9h-12h'' line intersects with the
horizontal ``0.6'' line at an rms of $560\,\mu$m, which means that,
statistically, in June between 09:00 and 12:00, the path length
fluctuations are smaller than $560\,\mu$m during 60\,\% of the time.

From the cumulative histograms we have obtained the fraction of
measurements that lie below $750\,\mu$m. These rms values are listed
in Table~\ref{tab:winter}. From May to August, between 09:00 and 12:00
and between 15:00 and 18:00, the mean of the fraction of measurements
which are smaller than $750\,\mu$m is $0.82\pm0.05$. This means that
in typical conditions on winter days, during 82\,\% of the time the
path length fluctuations are smaller than $750\,\mu$m and the
decorrelation at 86\,GHz on baselines of 150\,m or shorter is 10\,\%
or less. In the rest of the year at night, i.e., between 21:00 and
06:00, the mean is $0.87\pm0.07$. This means that the likelihood to
have seeing conditions suitable for millimetre observing is a little
higher outside the millimetre season at night, than it is in the
millimetre season during the day. The effect is particularly
pronounced in September, October, and April, whereas the nights of the
midsummer months December and January are slightly worse than the
average millimetre season days.

\begin{table*}[h]
\begin{center}
\begin{tabular}{lcccccccc}
\hline 
Month        & 00:00\,h - & 03:00\,h - & 06:00\,h - & 09:00\,h - & 12:00\,h - & 15:00\,h - & 18:00\,h - & 21:00\,h - \\
             & 03:00\,h    & 06:00\,h    & 09:00\,h    & 12:00\,h    & 15:00\,h    & 18:00\,h    & 21:00\,h    & 00:00\,h    \\
\hline
January      & 0.81 &   0.88 &   0.79 &   0.37 &   0.24 &   0.35 &   0.73 &   0.72  \\
February     & 0.83 &   0.77 &   0.76 &   0.38 &   0.15 &   0.33 &   0.73 &   0.72  \\
March	     & 0.91 &   0.89 &   0.80 &   0.33 &   0.22 &   0.44 &   0.72 &   0.78  \\
April	     & 0.94 &   0.93 &   0.93 &   0.65 &   0.37 &   0.76 &   0.90 &   0.94  \\
May          & 0.97 &   1.00 &   0.99 &   0.82 &   0.53 &   0.90 &   0.99 &   0.98  \\
June         & 0.94 &   0.92 &   0.92 &   0.78 &   0.55 &   0.81 &   0.89 &   0.92  \\
July         & 0.92 &   0.94 &   0.94 &   0.82 &   0.56 &   0.84 &   0.95 &   0.94  \\
August       & 1.00 &   0.98 &   0.97 &   0.74 &   0.38 &   0.82 &   0.97 &   0.96  \\
September    & 0.91 &   0.93 &   0.84 &   0.48 &   0.29 &   0.74 &   0.90 &   0.96  \\
October	     & 0.92 &   0.88 &   0.72 &   0.31 &   0.40 &   0.71 &   0.89 &   0.93  \\
November     & 0.89 &   0.91 &   0.66 &   0.24 &   0.16 &   0.58 &   0.82 &   0.81  \\
December     & 0.87 &   0.88 &   0.61 &   0.20 &   0.20 &   0.42 &   0.77 &   0.83  \\
\hline
\end{tabular}
\caption{Fraction of time where the rms of the path length fluctuations
is smaller than $750\,\mu$m.}
\label{tab:winter}
\end{center}
\end{table*}

We note that if the measurements taken between 12:00\,h and 15:00\,h
in the millimetre season months were taken into account, then the
fraction of time during which the path length fluctuations in this
period are better than $750\,\mu$m decreased to $0.71\pm0.16$. Then,
nighttime observations outside the millimetre season appeared even
more reasonable.

\subsection{Other considerations affecting the scheduling}

The scheduling of observations is not only constrained by weather, but
is a complex process which is influenced by many factors.

\begin{itemize}

\item Array configurations required for centimetre and millimetre
observations tend to be different. Millimetre observations generally
require compact configurations with spacings no more than a few
hundred metres whereas centimetre observations generally require
extended arrays. Because the compact configurations are not of
interest to centimetre observers, it is generally more efficient to
schedule the compact configurations at a time when both the days and
nights can be used for millimetre observations.

\item The seeing is not the only limiting factor for millimetre
observations. Atmospheric opacity, a strong function of the
atmosphere's total content of water vapour, is generally higher
outside the winter, and reduces the sensitivity of millimetre
observations.

\item The elapsed time between 21:00 and 06:00 is only 9\,h.
This is not sufficient for a full synthesis observation unless the
array is in one of the hybrid configurations, which are not frequently
scheduled outside the millimetre season. Hence these times would not
be suitable for imaging observations of complex structures, but might
be for detection experiments or monitoring of flux densities.

\end{itemize}

Furthermore, the amount of data is still small, which makes our
analysis susceptible to unusual weather conditions. For example, May
in 2005 was unusually dry, and June to August 2005 experienced a once
in 10 year wet spell, which may make the winter appear worse than it
is on average.

\section{Statistics of the speed of the phase screen}

\begin{figure}[htpb]
\centering
\includegraphics[width=5cm, angle=270]{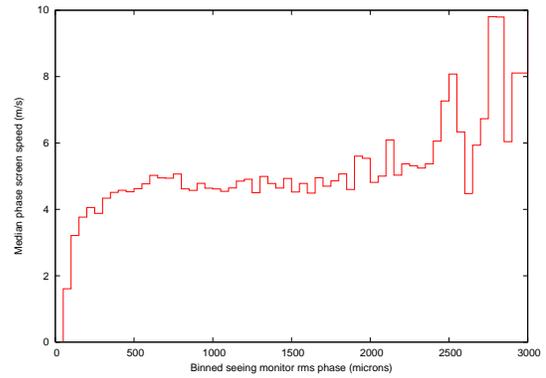}
\caption{The speed of the phase screen as a function of seeing monitor
rms phase. While the lowest and highest screen speeds are associated
with very low and high rms values, respectively, the screen speed is
otherwise largely independent of rms.}
\label{fig:screen_speed}
\end{figure}

Our analysis also yields a measure for the speed of the phase screen,
$v_{\rm s}$. In Figure~\ref{fig:screen_speed}, we have plotted the
median value of $v_{\rm s}$ as a function of the path length
fluctuations, which were binned into sections $50\,\mu{\rm m}$
wide. One can see that while the lowest screen speeds occur at the
lowest rms values, $v_{\rm s}$ is remarkably constant between
$500\,\mu{\rm m}$ and $2400\,\mu{\rm m}$, above which the $v_{\rm s}$
increases. The interpretation of this diagram is that the very best
rms values occur when the air above the observatory is extremely
still. The lowest rms phase occurs on still winter nights, when an
inversion layer has formed in the atmosphere: this is reflected in the
low screen speeds. On the other hand, the highest rms phase occur
during storms, and this seems to be reflected in high screen speeds.

\begin{figure}[htpb]
\centering
\includegraphics[width=7.5cm]{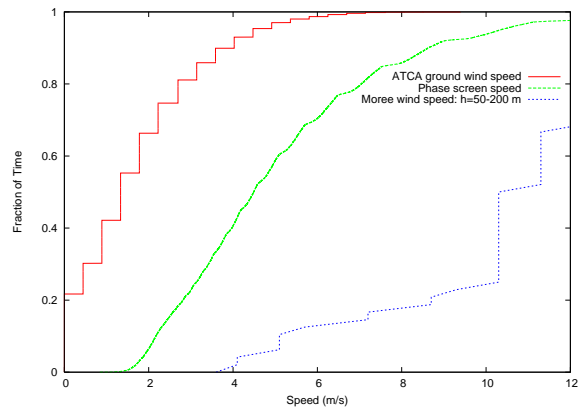}
\caption{A comparison of the cumulative histograms of the ATCA ground
wind speed, the ATCA phase screen speed and the speed of the winds at
Moree at an height of 50 to 200 metres above the ground. The steps in
some of the curves reflects quantisation in the measurements. The
jaggedness of the Moree data is accentuated by comparatively modest
number of measurements in this height band.}
\label{fig:wind_speed}
\end{figure}

It is interesting to compare the phase screen speed with the physical
wind speed both at the ground and at differing heights. We have
compared the phase screen speed with the median ground wind speed
simultaneously measured at the ATCA weather station. However we find
no clear relationship between the two. We have also compared them in a
statistical manner.  Figure~\ref{fig:wind_speed} gives a cumulative
histograms of the ATCA ground windspeed and the phase screen wind
speed inferred from the seeing monitor. On average, the phase screen
speed is greater than the ground wind speed.

We are not able to directly compare the phase screen speed with wind
levels at different heights above the ATCA. Instead we have obtained
wind readings with altitude measured by the Bureau of Meteorology's
radiosonde program at the township of Moree. Moree is approximately
100\,km to the north of the ATCA, and is in a topographically similar
setting. Like the ATCA, Moree lies on the western plains, with the
prevailing weather normally coming from the south-west or north-west.
Radiosonde measurements from 1987 to 2005 were used in the comparison.
Analysis of the radiosonde data shows that the wind is normally
comparatively low at the ground, that the wind increases significantly
as soon as the radiosonde balloons leave the ground, but then remains
fairly uniform with altitude up to about 2000\,m above the
ground. Comparison of the ground wind speed at Moree and the ATCA
shows that Moree is somewhat windier place. However the Moree ground
wind speeds are still lower than the ATCA phase screen speeds.

Figure~\ref{fig:wind_speed} also shows the cumulative histogram of the
wind speed measured by Moree radiosondes at heights between 50 and
200~m above the ground. This shows significantly higher speeds than
the ground wind or the phase screen speeds. As the phase screen speed
behaviour is more similar to the ground wind speed, this suggest that
the phase screen is close to the ground.

This conclusion is consistent with an excess amount of water vapour
seen in the Moree radiosonde measurements: whereas models of water
vapour content usually suggest that water vapour concentration falls
off exponentially with a scale height of $\sim$2~km, the Moree
radiosonde data show an extra component close to the ground.

\section{Statistics of the Kolmogorov exponent}

\begin{figure}[htpb]
\centering
\includegraphics[width=7.5cm]{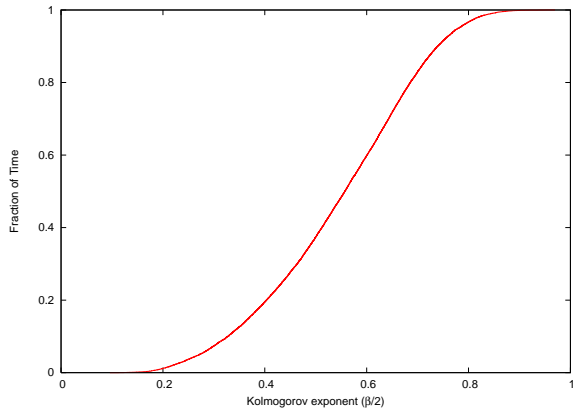}
\caption{A cumulative histogram of the Kolmogorov exponent, $\beta/2$.}
\label{fig:kol-cumulative}
\end{figure}

\begin{figure}[htpb]
\centering
\includegraphics[width=5cm, angle=270]{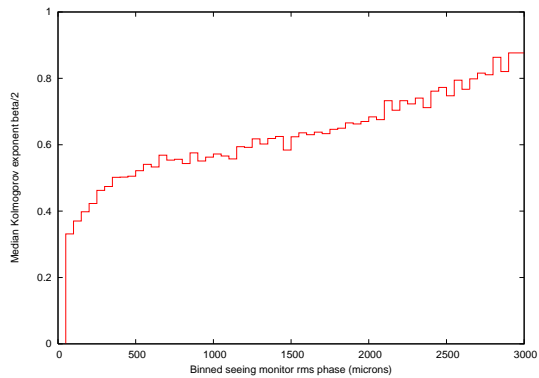}
\caption{The Kolmogorov exponent  as a function of seeing monitor
rms phase.}
\label{fig:kolmogorov}
\end{figure}

Figure~\ref{fig:kol-cumulative} and \ref{fig:kolmogorov} give the
Kolmogorov exponent, $\beta/2$ as a cumulative histogram and versus
rms phase. The latter figure gives the median binned in the same
fashion as the previous section.  The figures show that in the best
seeing conditions $\beta/2$ is 0.4 or even a bit lower, and so is near
the value of $1/3$ expected for 2D turbulence.  The exponent then
rises gently with increasing rms phase, and reaches a value of a bit
over 0.8 in the worst seeing conditions. A Kolmogorov exponent of
$5/6=0.83$ is expected in the case of 3D turbulence. The median
Kolmogorov exponent in our data is 0.56. This suggests that conditions
are in a transition domain between 3D and 2D turbulence, and implies
that the phase screen depth is comparable to the baseline length of
230\,m.

\section{Conclusion}

We have presented a description of the ATCA's seeing monitor along
with an analysis of 21 months of data. The main conclusions of this
analysis are that:

\begin{itemize}

\item the seeing conditions on summer nights are not substantially
different from winter days;

\item the turbulence seems to be concentrated quite near the ground;
and

\item the turbulence appears to be in transition between
being described by 2D and 3D models. This suggests the turbulent layer
is comparable to the seeing monitor baseline length.

\end{itemize}

The latter two points build a consistent picture of the turbulence
being substantially confined to a layer in the lower $\sim50 - 100$\,m
of the atmosphere where the interaction with the ground is likely to
be strongest.

\section*{Appendix A: Decorrelation of interferometer measurements due
to atmospheric seeing}

The expectation value for visibility measurements of a two-element
interferometer, $\langle V_{\rm m}\rangle$, is related to the true
visibilities, $V$, by

\begin{equation}
\langle V_{\rm m}\rangle = Ve^{-\sigma_\phi^2/2}
\label{eq:decorr}
\end{equation}

\noindent (\citealt{Thompson2001}, eq. 13.81), where $\sigma_\phi$ denotes
Gaussian-distributed random phase fluctuations. Random phase errors of
$30^\circ$ therefore lead to a degradation in observed visibility
amplitude to 87\,\% of the true value. Phase errors arising from water
vapour scale linearly with frequency, hence to convert seeing monitor
readings into a phase rms one can write

\begin{equation}
\sigma_\phi=2\pi\frac{L}{\lambda},
\label{eq:rms1}
\end{equation}

\noindent where $\lambda$ is the wavelength observed, and $L$ is the
rms of the path length fluctuations. Decorrelation is also a function
of the projected baseline length, $d$, and the structure of the
atmospheric turbulence. For short baselines, the tropospheric water
vapour fluctuations can be described as three-dimensional Kolomogorov
turbulence, where the amplitude of the turbulence scales with the size
of the turbulence cells as $\beta/2=5/6$, such that

\begin{equation}
\sigma_\phi=2\pi\frac{L}{\lambda}\left(\frac{d}{230\,{\rm m}}\right)^{5/6},
\label{eq:rms2}
\end{equation}

\noindent where the 230\,m is the baseline length of the seeing
monitor. Letting $V=1$ and solving for $L$ as a function of $\langle
V_{\rm m} \rangle$, one gets

\begin{equation}
L=\frac{\lambda}{2\pi}\sqrt{-2{\rm ln}\langle V\rangle}
\label{eq:rms3}
\end{equation}

\bibliography{refs}

\end{document}